\documentclass[10pt,a4paper]{article}

\begin{document}
\textwidth=135mm
\textheight=200mm

\begin{center} {\bfseries Renormalization Group Approach towards the
    QCD Phase Diagram}

  \vskip 5mm

  B.-J. Schaefer$^{1}$ and J. Wambach$^{2,3}$

  \vskip 5mm

  {\small {\it $^1$Institut f\"{u}r Physik, Karl-Franzens-Universit\"{a}t,
      Universit\"{a}tsplatz 5, A-8010 Graz, Austria}} \\ 
  {\small {\it $^2$ Institut f\"{u}r Kernphysik, TU Darmstadt,
      Schlo{\ss}gartenstra{\ss}e 9, D-64289 Darmstadt}}\\
  {\small {\it $^3$ Theory Division, GSI mbH, Planckstra{\ss}e 1,  D-64291 Darmstadt, Germany}}

\end{center}

\vskip 5mm

\centerline{\bf Abstract} 

The idea of the functional renormalization group and one-loop improved
renormalization group flows are reviewed. The associated flow
equations and nonperturbative approximations schemes for its solutions
are discussed. These techniques are then applied to the strong
interaction in the framework of an effective quark meson model which
is introduced in great detail. The renormalization group analysis of
the two flavor quark meson model is extended to finite temperature and
quark chemical potential which allows for an analysis of the chiral
phase diagram beyond the mean field approximation.
 
\vskip 10mm

\section{Introduction}
\label{sec:intro}

Quantum Chromodynamics (QCD) is the quantum field theory of strongly
interacting matter. Many of its vacuum features have been tested
experimentally over a wide range of momentum scales. At high momenta
or small distances the asymptotic freedom of non-Abelian gauge
theories can be used to apply perturbative methods for the computation
of physical observables. Due to the running of the strong gauge
coupling the situation becomes significantly more complicated at
smaller momentum scales or larger distances. Here perturbation theory
breaks down and nonperturbative methods are called for. There are no
analytical methods starting from first principles which allow to treat
QCD at larger distances, where the strong gauge coupling becomes large
and perturbation theory fails. The main reason for this difficulty is
that QCD describes qualitatively different physics at different length
or energy scales.

The situation is further complicated at finite temperature and/or
baryon density. For instance, at very high temperatures perturbative
calculations are plagued by serious infrared divergences. Furthermore,
it is generally expected that at high enough temperature and densities
hadronic matter attains a state in which chiral symmetry is almost
restored and its fundamental degrees of freedom, the quarks and
gluons, are no longer confined. The system undergoes a phase
transition from the ordinary hadronic phase to a chirally restored and
deconfined quark gluon plasma (QGP). Furthermore, recent theoretical
studies reveal an increasing richness in the structure of the phase
diagram of strongly interacting matter.

Therefore nonperturbative methods are indispensable to obtain
quantitatively reliable results because one has to deal with large
couplings and a perturbative expansion of physical observables fails.
One such nonperturbative method is given by the renormalization group
(RG). The RG method represents an efficient way to describe critical
phenomena and phase transitions. It can be used to characterize
universal and non-universal aspects of second-order as well as
first-order phase transitions and is very well adapted to reveal the
full phase diagram of strongly interacting matter. In the context of
the phase diagram the RG-method has been applied in
\cite{Schaefer2006a, Schaefer2005, Schaefer1999, Schaefer1997a} using
a two-flavor quark-meson model, which captures essential chiral
aspects of QCD. 

In the following we will introduce the concept of the effective
average action and its associated renormalization group equation.
Since it is clearly impossible to review here the many facets of the
renormalization group developments we will concentrate on the so
called proper-time renormalization group approach
\cite{Floreanini:1995aj, Liao1996}.

\section{Renormalization group methods}
\label{sec:RG}

The renormalization group deals with the effect of a scale change in a
theory. The central issue is the understanding of the macroscopic
physics at large distances or at low momenta in terms of the
underlying fundamental microscopic interaction. In order to understand
the evolution from the microscopic to the macroscopic scales one has
to consider the quantum or statistical fluctuations on all scales in
between. The general RG idea is to treat the fluctuations not all at
once but successively from scale to scale \cite{Wilson:1973jj}.

This RG idea combined with functional methods yields the so called
functional RG. By means of functional methods the computation of
generating functionals of correlation functions becomes feasible. All
important physical information is contained in the correlation
functions once the fluctuations have been integrated out. Instead of
evaluating correlation functions by averaging over all fluctuations at
once, only the change of the correlation functions induced by an
infinitesimal momentum shell of fluctuations is considered. This goes
along Wilson's philosophy of integrating out modes momentum shell by
momentum shell.

From a technical point of view this means one has to work with
functional differential equations - the so called RG or flow equations
- instead of functional path integrals which is usually the case in
standard quantum field theory. The differential structure of the RG
equations has a larger versatility and offers several advantages
compared to an integral formulation. It is analytically and
numerically better accessible and more stable. This is of great
interest in non-Abelian gauge theories such as QCD. During the
evolution from the microscopic to the macroscopic scales these
theories turn from a weak to a strong coupling becoming thus
nonperturbative at macroscopic scales. In this sense RG methods
provide an analytical and powerful tool to investigate nonperturbative
phenomena of quantum field theories and statistical physics.

There are various RG methods known in the literature
\cite{Litim:1998nf, Aoki:2000wm, Polonyi:2001se, Berges2002,
  Pawlowski:2005xe, Gies2006, bagnuls-2001-348}. One particular
formulation of RG flows is based on the concept of the effective
average action $\Gamma_k$, which is a simple generalization of the
standard effective action $\Gamma$, the generating functional of the
one-particle irreducible (1PI) Green functions
\cite{Wetterich:1992yh}. The generalization is achieved by
implementing an infrared cutoff scale $k$ in the functional integral
that defines the effective action. $\Gamma_k$ is then obtained by
integrating over all modes of the quantum fields with Euclidean
momenta larger than the infrared cutoff scale, i.e. $q^2>k^2$. In the
limit $k\to 0$, the infrared cutoff is removed and the effective
average action becomes the full quantum effective action $\Gamma$
containing all fluctuations. For any finite infrared cutoff $k$ the
integration of quantum fluctuations is only partially done. The
influence of modes with momenta $q^2 < k^2$ is not included.

On the other hand, in the limit $k\to \infty$ (or to some finite
ultraviolet cutoff $\Lambda$) the effective average action matches the
bare or classical action, which does not contain any fluctuations. 
Hence, the knowledge of the $k$-dependent $\Gamma_k$ allows to
interpolate from the bare action in the ultraviolet to the full
effective action in the infrared. As the scale $k$ is lowered more and
more quantum fluctuations are taken into account. As a consequence
$\Gamma_k$ can be viewed as a microscope with varying resolution
length scale $\sim 1/k$. It averages the pertinent fields over a
$d$-dimensional volume with size $1/k^d$ and permits to explore the
system on larger and larger length scales. This is similar to a
block-spin transformation on the sites of a coarse lattice where more
and more spin-blocks are averaged over.

The dependence of the effective average action $\Gamma_k$ on the scale
$k$ is governed by the flow equation
\begin{equation}\label{eq:ERG}
  \partial_t \Gamma_k [\phi]= \frac{ 1}{2} \mbox{Tr} \left(
\frac{\partial_t R_k}{\Gamma^{(2)}_k [\phi]+ R_k}\right)\ .
\end{equation} Here, $\Gamma^{(2)}_k [\phi]$ stands for the second
functional derivative with respect to some given field $\phi$. The
trace involves an $d$-dimensional integration over momenta (or
coordinates) as well as a summation over internal indices (e.g. 
flavor, color and/or Dirac indices). It is convenient to introduce the
logarithmic variable $t= \ln (k/\Lambda)$ which induces the
equivalence of the partial derivatives $\partial_t \equiv k
\partial/\partial k$.

The infrared cutoff function which regularizes the generating
functional is denoted by $R_k$ and depends on the IR scale $k$. This
regulator term, defined in momentum space, can be viewed as a
momentum-dependent mass term. The regulator function is not completely
arbitrary but has to satisfy certain conditions in order to provide an
IR and UV regularization for the flow equation. Since it screens the
IR modes in a mass-like fashion it should always be a positive
function regarding its arguments. Furthermore, it has to vanish for
$k\to 0$ removing the regulator from the flow. This condition ensures
that the full effective action is recovered in this limit. Finally, it
has to diverge for $k\to \infty$ (or for $k\to \Lambda$). This limit
ensures that the correct initial condition
$\lim\limits_{k\to \Lambda}\Gamma_k = S_\Lambda$ in the ultraviolet is
reached where $S_\Lambda$ denotes some bare action. This last
condition acts as a delta functional in the defining integral
representation of $\Gamma_k$ which is dominated by the stationary
point of the action. This limit justifies the use of a saddle-point
approximation which filters out the classical field configurations and
the bare action.

The flow equation (\ref{eq:ERG}) is an exact equation in the sense
that it can be derived from first principles and is often labeled as
Exact RG (ERG). It has a simple one-loop structure but is not of
one-loop order because it depends on the full field-dependent inverse
average propagator $\Gamma^{(2)}_k [\phi] + R_k$. This one-loop structure has
some advantages because technical complications due to overlapping
loop integrations do not arise.

The solution of the flow equation corresponds to a trajectory in the
theory space, the space of all possible action functionals which are
spanned by all possible invariant field operators. The starting point
of the trajectory in this space is the bare action $S_\Lambda$ and the
end point the full effective action $\Gamma$. As already mentioned,
the explicit form of the regulator function $R_k$ is not fixed
uniquely. Different choices for $R_k$ will lead to different
trajectories thus reflecting the RG scheme dependence of the flow. Of
course, the end point of all trajectories in the theory space is
independent of the choice of the explicit form of the regulator $R_k$.

General methods for the solution of the functional flow equation are
rare. Nevertheless, one can turn the flow equation into a coupled
system of non-linear ordinary partial differential equations for
infinitely many couplings by expanding the effective average action in
terms of invariants. In order to reduce the infinite system to a
numerically manageable size one needs to truncate the most general
form of $\Gamma_k$. This can be accomplished in various systematic
approximation schemes. One such approximation scheme is the 'operator
expansion' which constructs the effective action from operators with
increasing mass dimension.

An example for this type of truncation is the gradient expansion where
$\Gamma_k$ is expanded in numbers of derivatives. For a scalar field theory
with one real field one obtains
\begin{equation}
  \label{eq:gradexp}
  \Gamma_k[\phi] = \int d^dx \left\{ V_k(\phi) + \frac{ 1}{2} Z_k (\phi)
    (\partial_\mu \phi)^2 + {\cal O} ( \partial ^4 ) \right\}\ .
\end{equation}
The first (constant) term $V_k$ corresponds to the scalar effective
potential and the first correction includes the field-dependent wave
function renormalization $Z_k$. In general, it is not guaranteed that
the chosen truncation scheme converges. One has to control the
truncation error separately as for any expansion.

The flow of the effective average action, Eq.~(\ref{eq:ERG}), is not
the only exact one. It is closely related to other well-known exact
flows like e.g.~Wegner-Houghton flows \cite{Wegner1973} or
Callan-Symanzik flows \cite{Symanzik:1970rt, Callan:1970yg}.

Another class of RG flow equations can be obtained by approximations
to Exact RG flows. An example of such an approximation is the
so-called Proper-time RG (PTRG) flow which we want to investigate in
the following. Compared to the ERG flow the proper-time flow has a
numerically simpler and physically more intuitive representation and
yields in the lowest-order gradient expansion smooth and analytical
threshold functions.

\subsection{Proper-time RG flow}
\label{Sec:PTRG}

Proper-time flows in their original formulation are one-loop improved
RG flows \cite{Schaefer1999, Schaefer1997a, Liao1996}. They can be
obtained by an RG improvement in an one-loop flow equation
\cite{Schaefer2002, Schaefer:2000ce, Bohr2001, Papp2000}. Since the
PTRG flows do not depend linearly on the full field-dependent
propagator they cannot be exact flows \cite{litim-2002-66}.
Furthermore one can show that the (non-exact) PTRG flow starts to
deviate from standard perturbation theory already at two-loop order
\cite{litim-2002-65}.

Nevertheless, within a background field formalism a generalized PTRG
flow can be formulated which is exact. This is achieved by adding
further terms to the standard PTRG flow \cite{Litim:2002hj}. In the
following an ad hoc derivation of the standard one-loop improved PTRG
flow is presented. The discussion is restricted to an one-component
scalar theory with a general interaction. The generalization to other
theories is straightforward.

\subsubsection{Derivation of the standard PTRG flow}

The starting point for the derivation of an one-loop improved RG flow
is the one-loop effective action
\begin{equation}
  \Gamma_\Lambda^{\mbox{\scriptsize 1-loop}}[\phi] =
  S_{\mbox{\scriptsize class}} [\phi] + 
  \frac{ 1}{2} \mbox{Tr}
  \ln S_{\mbox{\scriptsize class}}^{(2)} [\phi]\ ,
\end{equation} 
where the trace denotes a sum over all momenta.
The classical action is labeled by
$S_{\mbox{\scriptsize class}} [\phi]$ and its second field-derivative
by
$S_{\mbox{\scriptsize class}}^{(2)}[\phi] ={\delta^2
  S_{\mbox{\scriptsize class}} [\phi]}/{\delta \phi \delta\phi}$. The
ill-defined logarithm is regularized by means of a Schwinger
proper-time representation
\begin{equation}\label{eq:oneloop}
  \Gamma_\Lambda^{\mbox{\scriptsize 1-loop}}[\phi] =
  S_{\mbox{\scriptsize class}} [\phi] - 
  \frac{ 1}{2}\int\limits_0^\infty \frac{ d\tau}{\tau}
  f_k(\tau)  \mbox{Tr}
  \exp ({-\tau S_{\mbox{\scriptsize class}}^{(2)} [\phi]})\ ,
\end{equation} 
where the regulator function $f_k (\tau)$ provides an IR cutoff $k$.
The lower (upper) limit of the proper-time integral has inverse mass
dimension two and regularizes the UV (IR) region of the theory. Since
the momenta and the proper-time variable $\tau$ are coupled, high
momenta correspond to small proper-time values and vice versa.
Taking the $k$-derivative of Eq.~(\ref{eq:oneloop}) the standard PTRG
flow equation is obtained as
\begin{equation} \label{eq:ptrg} \partial_t \Gamma_k [\phi] = - \frac{
    1}{2}\int\limits_0^\infty \frac{ d\tau}{\tau} \partial_t f_k(\tau)
  \mbox{Tr} \exp ( -\tau \Gamma^{(2)}_k [\phi])\ ,
\end{equation} 
where the classical action $S_{\mbox{\scriptsize class}}^{(2)}$ on the
right hand side has been replaced by the scale-dependent effective
action $\Gamma^{(2)}_k$. This represents the RG improvement
\cite{Liao1996}.

Similar to the ERG flow the regulator function $f_k$ has to satisfy
some conditions: Since the flow should start from the initial action
$\Gamma_\Lambda$ in the UV it is required that
$\lim\limits_{k\to\Lambda} f_k(\tau) =0$. The second condition
$\lim\limits_{\tau\to\infty} f_{k\neq0} (\tau) =0$ regularizes the IR
since the upper proper-time limit coincides with lower momenta.
Finally, at the end of the evolution, the condition
$\lim\limits_{k \to 0} f_k (\tau) =1$ reduces the proper-time
regularization to the usual Schwinger proper-time regularization.

One class of regulator functions $f^{(n)}_k (\tau)$ with $n\geq 0$,
which fulfill all these required conditions can be expanded in terms
of incomplete $\Gamma$-functions
and have the structure
\begin{equation}\label{eq:reg}
  \partial_t f^{(n)}_k (\tau ) = -2 \frac{ (\tau k^2)^{n+1}}{\Gamma
    (n+1)} e^{-\tau k^2}\ . 
\end{equation}
The $k^2$-term in the exponential acts as a mass term and controls the
IR behavior.
Inserting this basis set of regulators in Eq.~(\ref{eq:ptrg}) and
performing the proper-time integration yields the flow
\begin{equation}\label{eq:csflow}
  \partial_t \Gamma_k [\phi] = \mbox{Tr}  \left( \frac{k^2}{\Gamma^{(2)}_k [\phi]+
      k^2} \right)^{n+1} \ .  
\end{equation}

For $n>0$ the PTRG flow cannot be an exact flow since it does not
depend linearly on the full propagator. For $n=0$ the flow depends
linearly on the full propagator and one could conjecture that it is an
exact flow. In fact, it has the form of a Callan-Symanzik flow, but it
is not exact. Furthermore, the flow looks like an ERG flow with a mass
insertion $R_k = k^2$ but is not precisely an ERG flow. In contrast to
the ERG flow, the momentum integration is not regularized in the UV
because the condition $\lim\limits_{k\to \Lambda} R_k \to \infty$ is
not fulfilled in this case. All momenta contribute to the flow and an
additional UV renormalization would be required in contrast to the ERG
flow. In addition, for this flow the Wilsonian single momentum-shell
interpretation becomes questionable.

Nevertheless, the standard PTRG flow is a well-defined approximation
to a first-principles flow like the ERG flow \cite{Litim2006}. It is
an approximation to a background field ERG flow where additional terms
proportional to $\partial_t \Gamma^{(2)}_k$ are neglected and the
difference between fluctuation fields and background fields has been
omitted. With these additional terms the generalized PTRG flow becomes
exact. So far, applications of the PTRG flow are typically based on
flows of the form of Eq.~(\ref{eq:csflow}) and further approximations
thereof. The standard PTRG flow with the lowest-order gradient
expansion truncation has a simple polynomial momentum dependence which
will lead to simple flow equations. However, it is still an open issue
how these additional flow terms proportional to
$\partial_t \Gamma^{(2)}_k$ which make the flow exact affect the
results obtained with the standard PTRG flow. First results in this
respect for an $O(N)$ symmetric scalar theory indicate a minor
influence of these addition flow terms at least at criticality in
three dimensions \cite{Bohr2001, Litim2001}.

In the following the standard PTRG flow will be used to investigate
strongly interacting matter which is described by QCD.

\subsection{Applications}
\label{Sec:applications}

In QCD the quarks and gluons represent the microscopic degrees of
freedom, whereas the macroscopic degrees of freedom are the observed
color neutral particles like the mesons, baryons and/or glueballs.
Hence there must be a transition from the microscopic to the
macroscopic degrees of freedom. Therefore, in QCD the relevant degrees
of freedom change with the scale $k$ which is suitable for an RG
treatment. In order to apply RG techniques to QCD an initial starting
effective action has to be formulated.

When constructing effective models for these macroscopic degrees of
freedom one usually relies on the guiding symmetries of QCD because a
first-principle derivation from QCD is still missing. One important
symmetry of QCD is the local $SU(N_c)$ color invariance which is
related to confinement. This symmetry cannot be used here since the
observed hadronic spectrum consists of color blind states. We will
concentrate on the chiral dynamics of QCD and consider QCD with only
two light quark flavors. To a good approximation the masses of these
two flavors are small compared to the other quark flavors. For
vanishing current quark masses the classical QCD Lagrangian does not
couple left- and right-handed quarks. Omitting the axial anomaly and
baryon number conservation it exhibits a global chiral invariance
under the $SU_L(N_f)\times SU_R(N_f)$ symmetry group where $N_f$
denotes the number of quark flavors. In the observed hadron spectrum
only the vector-like subgroup $SU_V(N_f)$ is realized which implies a
spontaneous symmetry breaking of the chiral
$SU_L(N_f)\times SU_R(N_f)$ symmetry down to the $SU_V(N_f)$ symmetry.
This symmetry breaking predicts for $N_f=2$ the existence of three
light parity-odd Goldstone bosons, the three pions $\vec \pi$. Their
comparably small masses are a consequence of the explicit chiral
symmetry breaking due to the finite current quark masses.

At scales above $\sim 2$ GeV the dynamics of the relevant
degrees of freedom appear to be well described by perturbative QCD. At
somewhat lower scales quark condensates, bound states of quarks and
gluons emerge and confinement sets in. 

To each such non-perturbative phenomenon one can associate an
appropriate scale. Focusing on the physics of scalar and pseudoscalar
mesons and assuming that all other bound states are integrated out it
appears that all these scales are rather well separated from each
other. The compositeness scale $k_\phi$, where mesonic bound states
are formed due to the increasing strength of the strong interaction is
somewhere below $1$ GeV. The chiral symmetry breaking scale $k_\chi$
at which the chiral quark condensate develops a non-vanishing value
will be below the compositeness scale, typically around $500$ MeV. The
last scale where confinement sets in, is related to the Landau pole in
the perturbative evolution of the strong coupling constant and is of
the order of $\Lambda_{\mbox{\scriptsize QCD}} \sim 200$ MeV.

For scales between the ranges $k_\chi \leq k \leq k_\phi$ the most
relevant degrees of freedom are quarks and mesons and their dynamics
in this regime is dominated by the strong Yukawa coupling $g$ between
them. This picture legitimates the use of a quark meson model only if
one assumes that the dominant QCD effects are included in the meson
physics. Below the scale $k_\chi$ the strong coupling $\alpha_s$
increases further and quark degrees of freedom will confine. Getting
closer to $\Lambda_{\mbox{\scriptsize QCD}}$ it is not justified to
neglect those QCD effects which certainly go beyond the meson
dynamics. Of course, gluonic interactions are expected to be crucial
for an understanding of the confinement phenomenon. But due to the
increase of the constituent quark masses towards the IR the quarks
decoupled from the further evolution of the mesonic degrees of
freedom. As long as one is only interested in the dynamics of the
mesons one expects that the confinement on the mesonic evolution has
only little influence even for scales below
$\Lambda_{\mbox{\scriptsize QCD}}$. Hence there are good
prospects that the meson physics can be described by an effective
quark meson model~\cite{Tetradis:2003qa, Jungnickel1996,
  Jungnickel1998}. 

The Ansatz for the effective action at the compositeness scale
$k_\phi$ is given by the $N_f=2$ quark meson model
\begin{equation}
  \label{eq:qmmodel}
  \Gamma_{k=k_\phi} \!=\!\! \int\! d^4x \left\{ \bar{q} \,(i\partial\hspace{-2.1mm}/ + g
    (\sigma + i \gamma_5     \vec \tau \vec \pi ))\,q  +\frac 1 2
    (\partial_\mu \sigma)^2+     (\partial_\mu \vec \pi)^2 - U(\sigma,
    \vec \pi )\right\} 
\end{equation}
where the purely mesonic potential is defined as
\begin{eqnarray*}
  U(\sigma, \vec \pi ) &=& \frac \lambda 4 (\sigma^2+\vec \pi^2 -v^2)^2
  -c\sigma\ .
\end{eqnarray*} 
The quark fields are denoted by $q$ and $g$ is the Yukawa coupling
which describes the interaction strength of the isoscalar-scalar
$\sigma$ and the three isovector-pseudoscalar pion fields $\vec \pi$
with the quarks. The value for $g$ is usually fixed to the value of
the constituent quark masses $M_q$ which are of the order of $300$ MeV
in the vacuum. Here we have neglected the $SU(2)$-isospin violation
and consider averaged quark masses. The explicit symmetry breaking
parameter $c$ in the mesonic potential controls the value of the pion
mass. For $c=0$ the chiral limit is obtained and the pion mass vanish.
In the chiral limit the action is invariant under global chiral
$SU(2)_L\times SU(2)_R$ symmetry transformations. The last two
parameters of the model, $\lambda$ and $v^2$, are fixed to the mass of
the sigma field and the pion decay constant $f_\pi$.


This effective action at the compositeness scale $k_\phi$ emerges from
short distance QCD in basically two steps. Starting from QCD in the UV
one first computes an effective action which only involves quarks.
This corresponds to an integration over the gluonic degrees of freedom
in a quenched approximation. This will generate many effective
nonlocal four and higher quark vertices and a nontrivial momentum
dependence of the quark propagator. In the second step decreasing the
scale further these four and higher quark interactions will cause the
formation of mesonic bound states. Thus, at the compositeness scale
not only quarks but also composite fields, the mesons, are present and
interact with each other. The four quark interactions have been
replaced by mesonic fields. It is obvious that for scales below the
compositeness scale a description of strongly interacting matter in
terms of quark degrees of freedom alone would be rather inefficient.

The Ansatz for the effective action (\ref{eq:qmmodel}) incorporates a
truncation to four quark interactions. Higher quark interactions have
been suppressed at the compositeness scale. This leads to a purely
quadratic mesonic potential at $k_\phi$ with a positive mass term.
This implies that the chiral symmetry is restored at the compositeness
scale and the scalar expectation value $\langle \sigma\rangle$ will
vanish \cite{Schaefer1999}.

The effective action also corresponds to a gradient expansion to
lowest order (cf.~Eq.~(\ref{eq:gradexp}) where wave-function
renormalizations are neglected. It serves as the required initial
action for the RG evolution. The evolution is started at the
compositeness scale where the mesons are assumed to be formed due to
strong gluonic interactions. During the evolution towards the IR this
effective action for the quark meson model is used as a truncation for
an approximate solution of the RG flow. Instead of expanding the
mesonic potential in a polynomial form for the fields we will be more
general here and allow for arbitrary higher mesonic $O(4)$-symmetric
self-interactions in the potential. We will solve numerically the flow
equations for the full effective potential. But the running of the
Yukawa coupling will be neglected.

The dynamics in the beginning of the scale evolution just below the
compositeness scale is almost entirely driven by quark
fluctuations. These fluctuations rapidly drive the squared scalar mass
term in the action to negative values. This then immediately leads to
a potential minimum away from the origin such that the vacuum
expectation value $\langle \sigma \rangle$ becomes finite. This
happens at the chiral symmetry breaking scale $k_\chi < k_\phi$, not
far below $k_\phi$. The reason for this behavior lies in the
suppression of the meson contributions. All meson masses are much
larger than the constituent quark masses around these scales and are
therefore further suppressed during the evolution. Furthermore, the
quarks are strongly coupled to the mesons since the Yukawa coupling is
relatively large. Below $k_\chi$ the systems stays in the regime with
spontaneous chiral symmetry breaking. Around scales of the order of
the pion mass the evolution of the potential minimum stops. The reason
for this stability of the vacuum expectation value is that the quarks
acquire a relatively large constituent mass $M_q$. These heavy modes
will decouple from the further evolution once the scale drops below
$M_q$. The evolution is then essentially driven by the massless
Goldstone bosons in the chiral limit. Of course, for non-vanishing
pion masses the evolution of the model is effectively stopped around
scales $k \sim m_\pi$. Quarks below such scales appear to be no longer
too important for the further evolution of the mesonic system. Because
of confinement effects quarks should anyhow no longer be included for
scales below $\Lambda_{\mbox{\scriptsize QCD}}$. The final goal of
such an evolution is to extract phenomenological quantities like meson
masses, decay constants etc.~in the IR. This can be achieved in a
straightforward computation from the effective action
$\Gamma = \lim\limits_{k\to 0} \Gamma_k$.


So far we have considered the vacuum fluctuations that contribute to
the flow of the effective average action $\Gamma_k$. Using certain
phenomenological quantities in the IR as input we have fixed all model
parameters at the compositeness scale. The extension of the presented
RG formalism to a system in thermal equilibrium system with finite
net-baryon or net-quark number density is straightforward. In such
systems the effective average action plays the role of the grand
canonical potential $\Omega$ and depends also on the temperature $T$
and on one averaged quark chemical potential $\mu_q$ if again isospin
symmetry is assumed.

In the Imaginary-time or Matsubara formalism a finite temperature $T$
results in (anti-)periodic boundary conditions for (fermionic) bosonic
fields in the compact Euclidean time direction with radius $1/T$. This
leads to a replacement of the zeroth component of the momentum
integration with discrete (even) odd Matsubara frequencies for
(fermions) bosons. Thus, for finite temperature a four-dimensional
quantum field theory can be interpreted as a three-dimensional system
plus an infinite tower of Matsubara modes for each degree of freedom.
Since all Matsubara frequencies are proportional to the temperature
all massive Matsubara modes will decoupled from the dynamics of the
system at high temperature. One therefore expects an effective
three-dimensional theory with the bosonic zero modes as the only
relevant degrees of freedom. This leads to the phenomenon of
dimensional reduction.

The implementation of a finite quark chemical potential in the quark
meson model is also straightforward. One has to add a term
proportional to $i\mu_q \int d^4x \bar q \gamma_0 q$ to the Ansatz
(\ref{eq:qmmodel}) for the effective average action. 
This Ansatz together with the choice of a
half-integer regulator function $f_k^{(n)}$, see Eq.~(\ref{eq:reg}),
where the variable $n$ is replaced by $3/2$ is used as the input to derive
the proper-time flow equations. The choice of a half-integer power in
the regulator function corresponds to a three-dimensional momentum
cutoff and allows for an analytic evaluation of the Matsubara sums.
This is one advantage of using the proper-time flow with this
regulator compared to other RG flows. Finally, one obtains for the
scale-dependent grand canonical potential the proper-time flow
equation \cite{Schaefer2005, Schaefer2006a}
\begin{eqnarray}
  \label{eq:ptrgflow}
\partial_t \Omega_k(T,\mu) &=& \frac {k^5} {12\pi^2} 
\left[  \frac 3 {E_\pi} \coth \!\!\left(\frac {E_\pi}{2T} \right)\!+\!
    \frac 1  {E_\sigma} \coth \!\!\left(\frac
      {E_\sigma}{2T} \right) \right.\nonumber \\
  &&\hspace*{1cm}\left.- \frac {2 N_c N_f}{E_q}\left\{ \tanh\!\!  \left(\frac 
     {E_q -\mu_q} {2T}\right) +\tanh\!\! \left(\frac
     {E_q +\mu_q} {2T}\right)\right\}\right] ,
\end{eqnarray}
with the pion- $E_\pi = \sqrt{k^2 + 2 \Omega'_k}$, the $\sigma$-meson
$E_\sigma = \sqrt{k^2 + 2 \Omega'_k + 4\phi^2 \Omega''_k}$ and quark
energies $E_q = \sqrt{k^2 + g^2 \phi^2}$. The primed potential denotes
the $\phi^2$-derivative of the potential, i.e.,
$\Omega'_k := \partial \Omega_k / \partial \phi^2$ and correspondingly
the higher derivatives. The potential for this model depends on the
expectation value of the square of the chiral 4-component field
$\phi^2$ which coincides with ${\langle \sigma \rangle}^2$ since
${\langle {\vec \pi}}^2 \rangle=0$. The scale-dependent effective
meson masses are defined as
$m_{\sigma,k}^2 = 2 \Omega'_k + 4\phi^2 \Omega''_k$ and
$m_{\pi,k}^2 =2 \Omega'_k$ where the potential has to evaluated at the
global scale-dependent minimum $\phi^2 = \phi^2_{0}$. The dynamically
generated constituent quark mass $M_{q} = g \phi_0$ is proportional
to the minimum since the running of the Yukawa coupling is not
considered here.

The summations of the fermionic (bosonic) Matsubara sums yield
analytical functions which represent the corresponding mass threshold
functions for finite temperature and quark chemical potential. The
threshold functions are an important nonperturbative ingredient
generally appearing in flow equations. They control the smooth
decoupling of massive modes from the evolution once the IR cutoff
scale $k$ drops below the corresponding mass. The different degrees of
freedom contribute in an additive way to the flow. One recognizes the
three degenerate pion, the sigma and the quark/antiquark threshold
functions in the square brackets. The flow equation has an overall
scale factor $k^4$ which reflects the correct dimension of the
effective potential in $d=4$ dimensions. This can be seen explicitly
by rewriting all mass threshold functions in a dimensionless form. The
fermionic contributions enter with a negative sign due to the fermion
loop and have a degeneracy factor of $(2s+1) N_c N_f$ with $s=1/2$.
The quark chemical potential enters only in the quark/antiquark
threshold functions with the appropriate sign as it should be. It
influences the bosonic part of the flow equation only implicitly
through the meson masses. The flow for the vacuum can be deduced
immediately of the full flow (\ref{eq:ptrgflow}) by examining the
limits $T=0, \mu_q =0$ of the finite temperature and quark chemical
potential threshold functions. In this limit all hyperbolic functions
tend to one and only the vacuum threshold functions remain.

The flow equation (\ref{eq:ptrgflow}) constitutes a coupled, highly
non-linear, partial differential equation which can be integrated
numerically in principle in two ways: either one discretizes the
unknown potential $\Omega_k$ on a $\phi^2$-grid or expands the
potential in powers of $\phi^2$ around its minimum $\phi_0^2$. The
advantage of the potential expansion is that only a finite set of
coupled flow equations has to be solved, depending on the chosen
expansion order. In particular, this yields the beta functions for
the couplings of the potential. For each higher order of the
potential expansion, however, a new coupled beta function is introduced 
increasing the numerical effort drastically. A
further drawback is that the potential is only known around the
minimum $\phi^2_0$ once the system has been solved \cite{Papp2000,
  Schaefer2002}. This is different for the grid solution: Here, the
potential is not only known around the minimum but also for arbitrary
$\phi^2$. This is of importance, for example, in a first-order phase
transition where two degenerate minima of the potential emerge. In
this case the knowledge of all local minima is required to describe
the phase transition correctly. This is cumbersome in a potential
expansion, except for some simple potentials. With an additional
explicit symmetry breaking term in the potential every potential
minimum has always a finite value because the symmetry is never
restored exactly. A precise determination of the critical temperature
of a first-order transition is very difficult within an expansion
scheme around only one potential minimum.

Since we have opted for the grid solution, the field $\phi^2$ has been
discretized for a general potential term on a regular grid. Thus, for
each grid point a flow equation is obtained which leads finally to a
coupled closed system. As initial condition at the chiral symmetry
breaking scale $k_\chi$ we use a symmetric potential Ansatz. All model
parameters are fixed in such a way that they match the physical IR
vacuum quantities accordingly.

For zero chemical potential and for two massless quark flavors a
second-order phase transition is found in which the spontaneously
broken chiral symmetry is restored. The phase transition and fixed
point structure belongs to the $O(4)$-universality class with the
corresponding critical exponents. The value for the critical
temperature is a non-universal quantity and depends on the 
model parameters. For finite pion masses, i.e.~with an explicit
symmetry breaking term in the potential, the second-order phase
transition is washed out and turns into a smooth crossover. This
results in a shift of the ``pseudo-critical'' temperature defined as
the inflection point of the order parameter towards larger values.

For finite chemical potential, the $O(4)$-universal second-order phase
transition persists up to a tricritical point which is a critical
point where three phases coexist. The curvature of the second-order
transition line $T_c(\mu)$ has a negative slope. Thus, the location of
the tricritical point must be below the value of the critical
temperature at zero chemical potential. The tricritical point belongs
to a trivial Gaussian fixed point with mean-field critical exponents.
The precise location of this tricritical point is in general not known
and again depends on the model parameters~\cite{Brouzakis:2004se}.
Thus, the existence of this point, the shape of the transition line
and its universality class are predictions within the underlying quark
meson model. At higher chemical potential and smaller temperatures the
phase transition changes initially to a single first-order phase
transition. For smaller temperatures we observe a splitting of the
transition line and two phase transitions emerge. The left transition
line represents a first-order transition down to the $T=0$ axis. At
this transition the order parameter jumps not to zero but to a finite
value. The chiral symmetry remains spontaneously broken and is only
restored for higher chemical potentials which then produce the second
(right) transition line. At this right transition line we initially
find a second first-order transition where the order parameter jumps
to zero and chiral symmetry is restored. But for smaller temperatures
close the $\mu$-axis the order parameter tends smoothly to zero. It
seems that this transition is again of second-order. If this is true
we infer that there must be a second tricritical point in the phase
diagram in the chiral limit.

As the pions become massive the tricritical point turns into a
critical end point (CEP). For temperatures below the CEP a first-order
curved transition line is found which persists down to the $\mu$-axis.
For each value of the explicit symmetry breaking parameter (finite
pion mass), there is a corresponding CEP. In an extended,
three-dimensional ($T$,$\mu$,$m_\pi$) phase diagram these points
arrange into a critical line. The static critical behavior of this
line falls into the universality class of the Ising model in three
dimensions corresponding to the one-component scalar $\phi^4$-theory
in three dimensions. As the pions become massive no symmetry remains
which would require the order parameter to have more than one
component. On the other hand, in the chiral limit we have a chiral
$SU(2)_L\times SU(2)_R$ symmetry restoration which is isomorphic to
the $O(4)$ symmetry. In this case the order parameter should be made
up of four components, one sigma and three pions.

In the chiral limit below the splitting point, the right second-order
transition turns into a crossover for finite quark masses.
Analogously, the second tricritical point, if it exists, should turn
into a critical point. Some remnants of this critical point can indeed
be seen in the vacuum expectation value and in the behavior of the
meson masses. But a detailed analysis of this point is postponed to a
future work.

\section{Summary}
\label{Sec:sum}

In this lecture note we have presented an brief introduction to the
functional renormalization group methods with a focus on the flow
equation for the effective average action. We specialized and
discussed in great detail the standard proper-time flow and its
relation to the effective average action. In the second part we used
this proper-time RG to explore the QCD phase diagram within an linear
quark meson model for two quark flavors. This model captures essential
features of QCD such as the spontaneous breaking of chiral symmetry in
the vacuum and can therefore yield valuable insight into the critical
behavior associated with chiral symmetry.

\subsubsection*{Acknowledgment}

One of the authors (BJS) thanks the Organizers of the Helmholtz
International Summer School on ``Dense Matter in Heavy-Ion Collisions
and Astrophysics'' for the invitation. BJS would also like to thank
J.M. Pawlowski for numerous discussions and for a careful reading of
the manuscript.



\begin{thebibliography}{31}
\expandafter\ifx\csname natexlab\endcsname\relax\def\natexlab#1{#1}\fi
\expandafter\ifx\csname bibnamefont\endcsname\relax
  \def\bibnamefont#1{#1}\fi
\expandafter\ifx\csname bibfnamefont\endcsname\relax
  \def\bibfnamefont#1{#1}\fi
\expandafter\ifx\csname citenamefont\endcsname\relax
  \def\citenamefont#1{#1}\fi
\providecommand{\bibinfo}[2]{#2}
\providecommand{\eprint}[2][]{\url{#2}}

\bibitem{Schaefer2006a}
\bibinfo{author}{\bibfnamefont{B.-J.} \bibnamefont{Schaefer}} \bibnamefont{and}
  \bibinfo{author}{\bibfnamefont{J.}~\bibnamefont{Wambach}},
  \texttt{hep-ph/0603256}.

\bibitem{Schaefer2005}
\bibinfo{author}{\bibfnamefont{B.-J.} \bibnamefont{Schaefer}} \bibnamefont{and}
  \bibinfo{author}{\bibfnamefont{J.}~\bibnamefont{Wambach}},
  \bibinfo{journal}{Nucl. {P}hys.} \textbf{\bibinfo{volume}{A757}},
  \bibinfo{pages}{479} (\bibinfo{year}{2005}).

\bibitem{Schaefer1999}
\bibinfo{author}{\bibfnamefont{B.-J.} \bibnamefont{Schaefer}} \bibnamefont{and}
  \bibinfo{author}{\bibfnamefont{H.-J.} \bibnamefont{Pirner}},
  \bibinfo{journal}{Nucl. {P}hys.} \textbf{\bibinfo{volume}{A660}},
  \bibinfo{pages}{439} (\bibinfo{year}{1999}).

\bibitem{Schaefer1997a}
\bibinfo{author}{\bibfnamefont{B.-J.} \bibnamefont{Schaefer}} \bibnamefont{and}
  \bibinfo{author}{\bibfnamefont{H.-J.} \bibnamefont{Pirner}},
  \texttt{hep-ph/9712413}.

\bibitem{Floreanini:1995aj}
\bibinfo{author}{\bibfnamefont{R.}~\bibnamefont{Floreanini}} \bibnamefont{and}
  \bibinfo{author}{\bibfnamefont{R.}~\bibnamefont{Percacci}},
  \bibinfo{journal}{Phys. {L}ett.} \textbf{\bibinfo{volume}{B356}},
  \bibinfo{pages}{205} (\bibinfo{year}{1995}).

\bibitem{Liao1996}
\bibinfo{author}{\bibfnamefont{S.-B.} \bibnamefont{Liao}},
  \bibinfo{journal}{Phys. {R}ev.} \textbf{\bibinfo{volume}{D53}},
  \bibinfo{pages}{2020} (\bibinfo{year}{1996}).

\bibitem{Wilson:1973jj}
\bibinfo{author}{\bibfnamefont{K.~G.} \bibnamefont{Wilson}} \bibnamefont{and}
  \bibinfo{author}{\bibfnamefont{J.~B.} \bibnamefont{Kogut}},
  \bibinfo{journal}{Phys. {R}ept.} \textbf{\bibinfo{volume}{12}},
  \bibinfo{pages}{75} (\bibinfo{year}{1974}).

\bibitem{Litim:1998nf}
\bibinfo{author}{\bibfnamefont{D.~F.} \bibnamefont{Litim}} \bibnamefont{and}
  \bibinfo{author}{\bibfnamefont{J.~M.} \bibnamefont{Pawlowski}},
  \texttt{hep-th/9901063}.

\bibitem{Aoki:2000wm}
\bibinfo{author}{\bibfnamefont{K.}~\bibnamefont{Aoki}}, \bibinfo{journal}{Int.
  {J}. {M}od. {P}hys.} \textbf{\bibinfo{volume}{B14}}, \bibinfo{pages}{1249}
  (\bibinfo{year}{2000}).

\bibitem{Polonyi:2001se}
\bibinfo{author}{\bibfnamefont{J.}~\bibnamefont{Polonyi}},
  \bibinfo{journal}{Central {E}ur. {J}. {P}hys.} \textbf{\bibinfo{volume}{1}},
  \bibinfo{pages}{1} (\bibinfo{year}{2003}).

\bibitem{Berges2002}
\bibinfo{author}{\bibfnamefont{J.}~\bibnamefont{Berges}},
  \bibinfo{author}{\bibfnamefont{N.}~\bibnamefont{Tetradis}}, \bibnamefont{and}
  \bibinfo{author}{\bibfnamefont{C.}~\bibnamefont{Wetterich}},
  \bibinfo{journal}{Phys. {R}ept.} \textbf{\bibinfo{volume}{363}},
  \bibinfo{pages}{223} (\bibinfo{year}{2002}).

\bibitem{Pawlowski:2005xe}
\bibinfo{author}{\bibfnamefont{J.~M.} \bibnamefont{Pawlowski}},
  \texttt{hep-th/0512261}.

\bibitem{Gies2006}
\bibinfo{author}{\bibfnamefont{H.}~\bibnamefont{Gies}}, 
  \texttt{hep-ph/0611146}.

\bibitem{bagnuls-2001-348}
\bibinfo{author}{\bibfnamefont{C.}~\bibnamefont{Bagnuls}} \bibnamefont{and}
  \bibinfo{author}{\bibfnamefont{C.}~\bibnamefont{Bervillier}},
  \bibinfo{journal}{Phys. {R}ept.} \textbf{\bibinfo{volume}{348}},
  \bibinfo{pages}{91} (\bibinfo{year}{2001}).

\bibitem{Wetterich:1992yh}
\bibinfo{author}{\bibfnamefont{C.}~\bibnamefont{Wetterich}},
  \bibinfo{journal}{Phys. {L}ett.} \textbf{\bibinfo{volume}{B301}},
  \bibinfo{pages}{90} (\bibinfo{year}{1993}).

\bibitem{Wegner1973}
\bibinfo{author}{\bibfnamefont{F.~J.} \bibnamefont{Wegner}} \bibnamefont{and}
  \bibinfo{author}{\bibfnamefont{A.}~\bibnamefont{Houghton}},
  \bibinfo{journal}{Phys. {R}ev.} \textbf{\bibinfo{volume}{A8}},
  \bibinfo{pages}{401} (\bibinfo{year}{1973}).

\bibitem{Symanzik:1970rt}
\bibinfo{author}{\bibfnamefont{K.}~\bibnamefont{Symanzik}},
  \bibinfo{journal}{Commun. Math. Phys.} \textbf{\bibinfo{volume}{18}},
  \bibinfo{pages}{227} (\bibinfo{year}{1970}).

\bibitem{Callan:1970yg}
\bibinfo{author}{\bibfnamefont{J.}~\bibnamefont{Callan},
  \bibfnamefont{Curtis~G.}}, \bibinfo{journal}{Phys. {R}ev.}
  \textbf{\bibinfo{volume}{D2}}, \bibinfo{pages}{1541} (\bibinfo{year}{1970}).

\bibitem{Schaefer2002}
\bibinfo{author}{\bibfnamefont{B.-J.} \bibnamefont{Schaefer}},
  \bibinfo{author}{\bibfnamefont{O.}~\bibnamefont{Bohr}}, \bibnamefont{and}
  \bibinfo{author}{\bibfnamefont{J.}~\bibnamefont{Wambach}},
  \bibinfo{journal}{Phys. {R}ev.} \textbf{\bibinfo{volume}{D65}},
  \bibinfo{pages}{105008} (\bibinfo{year}{2002}).

\bibitem{Schaefer:2000ce}
\bibinfo{author}{\bibfnamefont{B.-J.} \bibnamefont{Schaefer}},
  \bibinfo{author}{\bibfnamefont{O.}~\bibnamefont{Bohr}}, \bibnamefont{and}
  \bibinfo{author}{\bibfnamefont{J.}~\bibnamefont{Wambach}},
  \bibinfo{journal}{Int. {J}. {M}od. {P}hys.} \textbf{\bibinfo{volume}{A16}},
  \bibinfo{pages}{2119} (\bibinfo{year}{2001}).

\bibitem{Bohr2001}
\bibinfo{author}{\bibfnamefont{O.}~\bibnamefont{Bohr}},
  \bibinfo{author}{\bibfnamefont{B.-J.} \bibnamefont{Schaefer}},
  \bibnamefont{and} \bibinfo{author}{\bibfnamefont{J.}~\bibnamefont{Wambach}},
  \bibinfo{journal}{Int. {J}. {M}od. {P}hys.} \textbf{\bibinfo{volume}{A16}},
  \bibinfo{pages}{3823} (\bibinfo{year}{2001}).

\bibitem{Papp2000}
\bibinfo{author}{\bibfnamefont{G.}~\bibnamefont{Papp}},
  \bibinfo{author}{\bibfnamefont{B.-J.} \bibnamefont{Schaefer}},
  \bibinfo{author}{\bibfnamefont{H.-J.} \bibnamefont{Pirner}},
  \bibnamefont{and} \bibinfo{author}{\bibfnamefont{J.}~\bibnamefont{Wambach}},
  \bibinfo{journal}{Phys. {R}ev.} \textbf{\bibinfo{volume}{D61}},
  \bibinfo{pages}{096002} (\bibinfo{year}{2000}).

\bibitem{litim-2002-66}
\bibinfo{author}{\bibfnamefont{D.~F.} \bibnamefont{Litim}} \bibnamefont{and}
  \bibinfo{author}{\bibfnamefont{J.~M.} \bibnamefont{Pawlowski}},
  \bibinfo{journal}{Phys. {R}ev.} \textbf{\bibinfo{volume}{D66}},
  \bibinfo{pages}{025030} (\bibinfo{year}{2002}{\natexlab{}}).

\bibitem{litim-2002-65}
\bibinfo{author}{\bibfnamefont{D.~F.} \bibnamefont{Litim}} \bibnamefont{and}
  \bibinfo{author}{\bibfnamefont{J.~M.} \bibnamefont{Pawlowski}},
  \bibinfo{journal}{Phys. {R}ev.} \textbf{\bibinfo{volume}{D65}},
  \bibinfo{pages}{081701} (\bibinfo{year}{2002}{\natexlab{}}).

\bibitem{Litim:2002hj}
\bibinfo{author}{\bibfnamefont{D.~F.} \bibnamefont{Litim}} \bibnamefont{and}
  \bibinfo{author}{\bibfnamefont{J.~M.} \bibnamefont{Pawlowski}},
  \bibinfo{journal}{Phys. {L}ett.} \textbf{\bibinfo{volume}{B546}},
  \bibinfo{pages}{279} (\bibinfo{year}{2002}{\natexlab{}}).

\bibitem{Litim2006}
\bibinfo{author}{\bibfnamefont{D.~F.} \bibnamefont{Litim}},
  \bibinfo{author}{\bibfnamefont{J.~M.} \bibnamefont{Pawlowski}},
  \bibnamefont{and} \bibinfo{author}{\bibfnamefont{L.}~\bibnamefont{Vergara}},
  \texttt{hep-th/0602140}.

\bibitem{Litim2001}
\bibinfo{author}{\bibfnamefont{D.~F.} \bibnamefont{Litim}} \bibnamefont{and}
  \bibinfo{author}{\bibfnamefont{J.~M.} \bibnamefont{Pawlowski}},
  \bibinfo{journal}{Phys. {L}ett.} \textbf{\bibinfo{volume}{B516}},
  \bibinfo{pages}{197} (\bibinfo{year}{2001}).

\bibitem{Tetradis:2003qa}
\bibinfo{author}{\bibfnamefont{N.}~\bibnamefont{Tetradis}},
  \bibinfo{journal}{Nucl. {P}hys.} \textbf{\bibinfo{volume}{A726}},
  \bibinfo{pages}{93} (\bibinfo{year}{2003}).

\bibitem{Jungnickel1996}
\bibinfo{author}{\bibfnamefont{D.-U.} \bibnamefont{Jungnickel}}
  \bibnamefont{and}
  \bibinfo{author}{\bibfnamefont{C.}~\bibnamefont{Wetterich}},
  \bibinfo{journal}{Phys. {R}ev.} \textbf{\bibinfo{volume}{D53}},
  \bibinfo{pages}{5142} (\bibinfo{year}{1996}).

\bibitem{Jungnickel1998}
\bibinfo{author}{\bibfnamefont{D.-U.} \bibnamefont{Jungnickel}}
  \bibnamefont{and}
  \bibinfo{author}{\bibfnamefont{C.}~\bibnamefont{Wetterich}},
  \bibinfo{journal}{Eur. {P}hys. {J}our.} \textbf{\bibinfo{volume}{C2}},
  \bibinfo{pages}{557} (\bibinfo{year}{1998}).

\bibitem{Brouzakis:2004se}
\bibinfo{author}{\bibfnamefont{N.}~\bibnamefont{Brouzakis}} \bibnamefont{and}
  \bibinfo{author}{\bibfnamefont{N.}~\bibnamefont{Tetradis}},
  \bibinfo{journal}{Nucl. Phys.} \textbf{\bibinfo{volume}{A742}},
  \bibinfo{pages}{144} (\bibinfo{year}{2004}).

\end{thebibliography}
\end{document}